\documentclass[12pts, draft, onecolumn]{IEEEtran}
\usepackage{cite}
\usepackage{amsmath,amssymb,amsfonts}
\usepackage{algorithmic}
\usepackage{graphicx}
\usepackage{textcomp}
\usepackage{soul}
\usepackage[utf8]{inputenc}
\usepackage[T1]{fontenc}
\usepackage{booktabs,
            makecell, 
            multirow,
            tabularx} 

\usepackage{siunitx}  

\def\BibTeX{{\rm B\kern-.05em{\sc i\kern-.025em b}\kern-.08em
    T\kern-.1667em\lower.7ex\hbox{E}\kern-.125emX}}
\begin{document}
\title{{\fontsize{24}{26}\selectfont{Communication\rule{29.9pc}{0.5pt}}}\break\fontsize{16}{18}\selectfont
Closed-Form Directivity Expression for Arbitrary Volumetric Antenna Arrays}
\author{Bruno Felipe Costa, Taufik Abrão, \IEEEmembership{Senior, Member IEEE}
\thanks{Submission on May 5th, 2018. This work was supported in part by The National Council for Scientific and Technological Development (CNPq) of Brazil under Grants 304066/2015-0, and in part by CAPES - Coordenação de Aperfeiçoamento de Pessoal de Nível Superior, Brazil (scholarship),  and by the Londrina State University - Paraná State Government (UEL).}
\thanks{B. F. Costa and T. Abrão are with Universidade Estadual de Londrina (UEL), Londrina, Paraná, Brazil, taufik@uel.br, \,\,\, bruno.uel.felipe@gmail.com}
}

\maketitle

\begin{abstract}
It is proposed a closed-form expression of directivity for an arbitrary volumetric antenna arrays using a general element pattern expression of type $\sin^u{(\theta)}\cos^v{(\theta)}$, 
with $v > -\frac{1}{2}$ and $u > -1$, and $u, v \in \mathbb{Z}$. Variations of this expression for different values of $v$ and $u$ are analyzed from the analytical and numerical perspectives. The parameters found in the closed-form expression are related to the order $v$ and $u$ of the element patterns, the rectangular spatial coordinate of each antenna element, the magnitude and phase excitation coefficients (complex excitation) of all elements, and the desired angle in spherical coordinates $(\theta_0, \phi_0)$. The expression found in this work has been validated by numerical results, considering distinct configuration scenarios.
\end{abstract}


\section{Introduction}

Different types of antennas array have been considered in the transmission of multiple-input multiple-output (MIMO) systems given {their} capacity {to} provide directive beams with high antenna gains. In \cite{Abouda2006}, the effects of different antenna array geometries, such as uniform linear, uniform circular, uniform rectangular and uniform cubic arrays are investigated  in different MIMO channel {configurations}. Moreover, large-scale antenna arrays have been studied for a variety of applications{; for instance,} with the increase of the {number} of antennas, the massive MIMO (M-MIMO) condition {can be} attained, which is a promising solution to deal with the actual high-data rate {trend} scenarios. Indeed, M-MIMO technology has been identified as one of the promising technology to achieve 5G network requirements \cite{Agiwal2016} since {a massive number of antennas at the base station (BS) provides a considerable} improvement on the directivity and antenna gains compared to {conventional} MIMO systems. Deploying a BS with a large number of antennas can serve a large number of users at the same time and frequency, saving the scarce time-frequency resources \cite{Mbeutcha2016}. By increasing the number of BS antennas, the network capacity is improved substantially in multi-user scenarios, {as well as} throughput, link reliability, spectral and transmit-energy efficiency performance, while guaranteeing simpler signal processing.

It can be {verified} that 3D configuration of antenna elements (volumetric arrays) have become suitable for M-MIMO instead of 1D configuration (linear arrays) due to the constraint in array aperture \cite{Adnan2017}. However, directivity expression for such arrays are no easily found. High directive arrays provide substantial energy efficiency gains since the intensity of the signal is increased in the desired angles. Commonly, the radiation pattern of a single element is relatively large{, providing} small values of directivity. Antenna arrays produce gains in the directivity without increasing the size of single elements. Usually, the general array properties can be controlled and optimized by adjusting the number of elements, the {antenna} spacing between them, excitation coefficients, {phase of each element}, the geometrical configuration of the overall arrays (linear, circular, elliptical, and so on){, as well as} the relative pattern of the individual elements \cite{balanis2012}. 

In \cite{Nuttall2001} approximations for linear, planar and volumetric aperture and arrays directivity are derived, and numerical methods are applicable for such computations; however, the associated computational cost increases with the array geometry dimension and number of elements. Hence, in the last decades, several attempts have been made to evaluate directivity analytically. Closed-form expressions of directivity have been proposed in\cite{Haupt2010} for linear arrays equipped with isotropic antenna elements; {while} symmetric linear antenna arrays with short dipole and isotropic element patterns directivity {were} discussed in \cite{Werner1999}. Moreover, \cite{Singaraju1975} analyses single ring circular and elliptical arrays equipped with isotropic elements, while rectangular planar arrays with progressive phasing are treated in \cite{Bhattacharyya2010}. {Generalized expressions for} directivity {of} planar arrays have been investigated in \cite{Das_2013a} and \cite{Das_2013b}. Recently, a generalized expression of directivity for volumetric antennas using infinite series was presented in \cite{Das_2017}.

\noindent \textbf{\textit{Contribution}}. In this {\it communication}, a new and generalized expression for the directivity considering {also} a generalized element pattern and using arbitrary volumetric arrays is derived and corroborated by numerical analysis. 

This work is organized as follows. Section \ref{sec:pattern} describes the radiation pattern of an antenna array considering generalized element factors of type $\sin^u(\theta) \cos^v(\theta)$. In Section \ref{sec:voldirectivity}, a closed-form expression for volumetric antenna array directivity is derived using generalized element patterns. The found expressions for the directivity are validated by numerical analysis in Section \ref{sec:valid}, considering different scenarios of interest. Main conclusions are offered in Section \ref{sec:concl}.

\section{Radiation Pattern of an Antenna Array}\label{sec:pattern}

The total field of an array according to pattern multiplication principle can be formed by multiplying the field of a single element $(\Upsilon_e (\theta , \phi) )$ and the array factor $(\Upsilon_a (\theta , \phi) )$ {by the} selected direction. Hence, the radiation pattern of an antenna array can be written as:
\begin{equation}
\Upsilon (\theta , \phi)  = \Upsilon_e (\theta , \phi) \Upsilon_a (\theta , \phi)
\label{eq:1}
\end{equation} 
The element factor (EF), also {known} as element pattern, is the radiation pattern of a single-element antenna{;} the model selected must resemble the pattern of
a real antenna. Using spherical coordinates{,} the {resulted} EF {expression} is periodic in both $\theta$ and $\phi$-planes. Therefore, the radiated power of an arbitrary real antenna can be approximate using Fourier analysis which {results} in a linear combination of powers of cosines and sines. {Hence, in} this work the element factors will be considered {as}:
\begin{equation}
\Upsilon_e (\theta)  = \sin^u{(\theta)}\cos^v{(\theta)}
\label{EF}
\end{equation}
Indeed,  {we aim to find} a generalized directivity expression considering different values of {exponents} $u$ and $v$.

The array factor quantifies the effect of combining radiating elements in an array without {consider} the specific radiation pattern taken into account; this factor {has a dependency in terms of} position, relative phase and relative amplitude of each antenna element and can be written as{:}
\begin{equation}
\Upsilon_a (\theta,\phi) = \sum \limits_{n=1}^N A_ne^{j(\alpha_n + k \boldsymbol{r}_n . \boldsymbol{a}_r)}
\label{arraygeral}
\end{equation}
where $N$ is the number of elements, $A_n$ {is} the relative amplitude of $n$-th element, $\alpha_n$ {is} the relative phase of $n$-th element, $\boldsymbol{r}_n$ {is} the position vector of $n$-th element, $k$ {is} the wave number, and $\boldsymbol{a}_r$ {is} the unit vector of observation point in spherical coordinates, given by:
\begin{equation}
\boldsymbol{a}_r = \sin (\theta) \cos(\phi) \hat{i} + \sin (\theta) \sin(\phi)  \hat{j} + \cos (\theta) \hat{k}
\label{ar}
\end{equation}
The position vector utilized in this work is given by:

\begin{equation}
\boldsymbol{r}_n = r_{x_{n}} \hat{i} + r_{y_{n}} \hat{j} + r_{z_{n}} \hat{k}
\label{position}
\end{equation}

The array factor can be interpreted as the radiation pattern by replacing the actual elements by the isotropic (point) sources, \textit{i.e.},  $\Upsilon_e(\theta,\phi) = 1$.

\section{Closed-Form Expression for the Volumetric Array Directivity}\label{sec:voldirectivity}

When {the antenna does} not radiate power isotropically, \textit{i.e.}, equally in all directions, {the resulted directivity} can be desirable depending on the application, irradiating more signal power {in specific directions.} This feature is the result of constructive and destructive interference from radiators, {\it i.e.}, the elements of an antenna array. 

The directivity determination evaluates the power density in a designated direction to the average power density of irradiation in all directions. Antenna directivity can be expressed as the ratio between the radiation intensity  in the desired angle ($\theta_0 , \phi_0$) and the sum of the radiation intensity in all the {other directions}:
\begin{equation}
\mathcal{D}(\theta_0 , \phi_0) = \frac{|\Upsilon (\theta_0 , \phi_0)|^2 }{\frac{1}{4\pi} \int \limits_{0}^{2\pi} \int \limits_{0}^{\pi} |\Upsilon(\theta , \phi) |^2\sin{(\theta)} d\theta d\phi}
\label{eq:direnorm} 
\end{equation}

The numerator of \eqref{eq:direnorm}  is well-defined for a desired angle $(\theta_0,\phi_0)$ using \eqref{eq:1}; hence, solving the normalization integral of the denominator of \eqref{eq:direnorm}, a closed-form expression for the directivity {can be} obtained.
\begin{equation}
\mathcal{T} = \frac{1}{4\pi} \int \limits_{0}^{2\pi} \int \limits_{0}^{\pi} |\Upsilon(\theta , \phi)|^2 \sin{(\theta)} d\theta d\phi
\label{eq:int_} 
\end{equation}
The radiation intensity can be written considering \eqref{EF} and using \eqref{arraygeral}, \eqref{ar} and \eqref{position} as:
\begin{align}
\begin{split}
|\Upsilon(\theta,\phi)|^2 =& \sin^{2u}{(\theta)}\cos^{2v}{(\theta)} \Biggr [\sum \limits_{n=1}^N A_n^2 \\
&+ 2\sum\limits_{\substack{n,m=1 \\ m\neq n \\ n > m}}^N A_n A_m \cos{\left[\Omega_{xyz}(\theta,\phi,n,m)\right]} \Biggr]
\end{split}
\end{align}
where
\begin{align}
&\Omega_{xyz}(\theta,\phi,n,m) = k \, [(r_{x_n} - r_{x_m})\sin{\theta} \cos{\phi}\,\, \\
& + (r_{y_n} - r_{y_m}) \sin{\theta} \sin{\phi} + (r_{z_n} - r_{z_m}) \cos{\theta} ] + (\alpha_n-\alpha_m), \notag
\end{align}
in which we can rewrite as:
\begin{align}
\Omega_{xyz}(\theta,\phi,n,m)=& k\, \left (x_{mn} \sin{\theta} \cos{\phi} + y_{mn} \sin{\theta} \sin{\phi} \,\, \,\, \right. \notag\\ 
& \left. + \,\,z_{mn} \cos{\theta} \right)  + \alpha_{mn} 
\end{align}
where:
\begin{equation}
\begin{array}{lll}
 x_{mn} = k \left(r_{x_n} - r_{x_m}\right); \quad 
y_{mn} = k \left(r_{y_n} - r_{y_m}\right); \\
z_{mn} = k \left(r_{z_n} - r_{z_m}\right); \quad \,\,
\alpha_{mn} = \alpha_n - \alpha_{m}
\end{array}
\end{equation}
Therefore, the first integral in \eqref{eq:int_} can be expressed as:
\begin{align}\label{integral_appendix1} 
\mathcal{F} & = \,\, \frac{1}{4\pi} \int \limits_{0}^{2\pi} |\Upsilon(\theta , \phi)|^2  d\phi = \frac{1}{4\pi} \sin^{2u}{(\theta)}\cos^{2v}{(\theta)} \,\,\,\times\\
\hspace{-5mm} & \hspace{-5mm} \Biggr [ \int \limits_{0}^{2\pi} \sum \limits_{n=1}^N A_n^2 d\phi + 2 \int \limits_{0}^{2\pi} \sum\limits_{\substack{n,m=1 \\ m\neq n \\ n > m}}^N A_n A_m \cos{\left[\Omega_{xyz}(\theta,\phi,n,m)\right]}  \,\, d\phi \Biggr]\notag 
\end{align}
These two integrals is calculated in the Appendix \ref{app:A}, resulting:
\begin{align}
	\begin{split}
\mathcal{F} =& \frac{1}{2} \sin^{2u}{(\theta)}\cos^{2v}{(\theta)} \Biggr [\sum \limits_{n=1}^N A_n^2 \\
&+ 2\sum\limits_{\substack{n,m=1 \\ m\neq n \\ n > m}}^N J_0 \left (\gamma_{mn}\right )  \, A_n A_m \cos{(z_{mn} \cos{\theta} + \alpha_{mn} )} \Biggr]
	\end{split}
\end{align}
where $\gamma_{mn} =\sin{\theta} \sqrt{x_{mn}^2 + y_{mn}^2}$, and $J_0(\cdot)$ is the Bessel function of the first kind. Hence, the directivity becomes:
\begin{equation}
\mathcal{D}(\theta_0 , \phi_0) = \frac{|\Upsilon(\theta_0,\phi_0)|^2}{\mathcal{T}} = \frac{|\Upsilon(\theta_0,\phi_0)|^2}{\int \limits_0^\pi \mathcal{F} \sin{(\theta)} d\theta}
\label{integral_appendix2}
\end{equation}

Solving the integral in \eqref{integral_appendix2} (see the Appendix \ref{app:B}), the directivity for an arbitrary volumetric antenna array results in \eqref{equation_geral}, which represents a new analytical closed-form generalized expression for the antenna array directivity.
\begin{figure*}[!t]
\normalsize
\begin{align}
&\mathcal{D}(\theta_0 , \phi_0) = \frac{\sin^{2u}{(\theta_0)}\cos^{2v}{(\theta_0)} \Biggr [\sum \limits_{n=1}^N A_n^2 + \sum\limits_{\substack{n,m=1 \\ m\neq n \\ n > m}}^N A_n A_m \cos{\left[\Omega_{xyz}(\theta_0,\phi_0,n,m)\right]} \Biggr]}{\sum \limits_{n=1}^N A_n^2 \left[ \frac{1}{8}((-1)^{2v}+1)\mathcal{B}(u+1,v+\frac{1}{2}) \right] + 2 \, (-1)^{(v+2u)} \sum\limits_{\substack{n,m=1 \\ m\neq n \\ n > m}}^N \sum \limits_{\kappa = 0}^u A_n A_m \binom{u}{\kappa} \cos{(\alpha_{mn})}\frac{\partial^{2(v+u-\kappa)}}{\partial z_{mn}^{2(v+u-\kappa)}} \left [\frac{\sin (\sqrt{\beta^2+z_{mn}^2})}{\sqrt{\beta^2+z_{mn}^2}} \right ] } \notag\\
&\text{where:}  \qquad \Omega_{xyz}(\theta_0,\phi_0,n,m) = k \, [(r_{x_n} - r_{x_m})\sin{\theta_0} \cos{\phi_0} + (r_{y_n} - r_{y_m}) \sin{\theta_0} \sin{\phi_0} + (r_{z_n} - r_{z_m}) \cos{\theta_0} ] + (\alpha_n-\alpha_m) \notag\\
&\beta = k \sqrt{(r_{x_{n}}-r_{x_{m}})^2 + (r_{y_{n}}-r_{y_{m}})^2};  \qquad  z_{mn} = k (r_{z_{n}}-r_{z_{m}}); \qquad u > -1;\qquad v> -\frac{1}{2};  \quad \text{and} \quad  v,u \in \mathbb{Z}
\label{equation_geral}
\end{align}
\hrulefill
\vspace*{4pt}
\end{figure*}

Since the proposed directivity expression does not use any approximation, the result {found} is expected to be  exact. Moreover, such expression {does} not need to estimate the pattern and integrate over the entire beam area; therefore, computational efforts and time consumption are mitigated.

\section{Validation}\label{sec:valid}
The validation of the proposed directivity closed-form expression in \eqref{equation_geral} was conducted through extensive numerical integration and analysis. In the sequel{, the closed-form directivity is evaluated  considering} illustrative scenarios of interest.

\subsection{Scenarios}
All simulations and results follow the parameter setup described in Table \ref{table1}, which is the same utilized in \cite[Tab. I, Set 2]{Das_2017}. The direction selected to evaluate the directivity was $\theta_0 = \ang{101.44}$ and $\phi_0 = \ang{267.75}$, which is the angle that maximize the directivity for omnidirectional sources ($u=0 \,|\, v=0$). The directivity was calculated for different values of EF using \eqref{equation_geral} and compared with numerical integration. {Adopting omnidirectional sources, it is possible to compare and corroborate the results obtained with those in \cite{Das_2017}.}

\vspace{-2mm}

\begin{table}[!htbp]
\centering
\caption{Parameter Setup used in the Numerical Simulations.}\label{table1} 
\begin{tabular}{ccc}
 \hline 
    \textbf{Element Location}   &  \multicolumn{2}{c}{\textbf{Complex Excitation}}  \\ \hline  
$ \left [r_{x_n} , r_{y_n}  , r_{z_n} \right ]\cdot \,\lambda $ & $A_n$ (normalized) & $\alpha_n$ (degree)  \\ \hline \hline 
    $ \left [0.00 , 0.00  , 0.00 \right ] $  & 0.84 & -4.28             \\ 
    $ \left [2.71 , 1.22  , 1.06\right ] $  & 0.93 & -121.36            \\ 
    $ \left [1.03 , 4.77  , 4.40\right ] $  & 0.13 & -58.39             \\
    $ \left [5.31 , 3.61  , 2.45\right ] $  & 0.94 & 74.06              \\
    $ \left [0.52 , 2.25  , 1.71\right ] $  & 0.60 & -112.04            \\
    $ \left [4.37 , 3.07  , 2.52\right ] $  & 0.10 & 169.73             \\
    $ \left [4.59  , 2.51  , 0.98\right ] $ & 0.29 & -59.63             \\
    $ \left [4.84 , 0.88  , 1.16\right ] $  & 0.56 & 143.29             \\
    $ \left [0.92 , 1.70  , 5.21\right ] $  & 0.99 & -94.67             \\
    $ \left [2.50 , 1.12  , 5.28\right ] $  & 1.00 & 124.53             \\ \hline 
 \multicolumn{3}{l}{$\lambda$: carrier wavelength}
 \end{tabular}
 \end{table}

\noindent The \textit{elements factors} considered include four configurations:
\begin{equation}\label{eq:ef}
\Upsilon_e (\theta)  = \sin^u{(\theta)}\cos^v{(\theta)} \equiv
\begin{cases} 
1  \, \, \quad \quad \quad \quad \quad \quad v=0 \, | \, u = 0 \\ 
\sin{(\theta)} \, \, \quad \quad \quad \quad v=0 \, | \, u = 1 \\ 
\cos{(\theta)} \, \quad \quad \quad \quad v=1 \, | \, u = 0 \\
\sin{(\theta)}\cos{(\theta)} \, \quad v=1 \, | \, u = 1 \\
\end{cases}
\end{equation}
The closed-form expression of directivity in terms of $\mathcal{T}$, as in \eqref{integral_appendix2}, for the four cases of \eqref{eq:ef}, is discussed in the sequel.

\noindent {\bf 1.} {\it Directivity for $v=0 \, | \, u = 0$}. For such values of {exponents},  the normalization integral, using \eqref{equation_geral}, results:
\begin{align}
\mathcal{T}_{1}  = \sum \limits_{n=1}^N A_n^2 + 2 \sum\limits_{\substack{n,m=1 \\ m\neq n \\ n > m}}^N A_n A_m \cos{(\alpha_{mn})} \frac{\sin (\sqrt{\beta^2+z_{mn}^2})}{\sqrt{\beta^2+z_{mn}^2}}
\end{align}
\noindent {\bf 2.} {\it Directivity for $v=1 \, | \, u = 0$}. For these values of {exponents}, the normalization integral is obtained as \eqref{wp2}. 
\begin{figure*}[!t]
\normalsize
\begin{align}
\begin{split}
\mathcal{T}_{2} = \sum \limits_{n=1}^N \frac{A_n^2}{3} - 2 \sum\limits_{\substack{n,m=1 \\ m\neq n \\ n > m}}^N A_n A_m \cos{(\alpha_{mn})} \left [\frac{(\beta^2-2z_{mn}^2)\cos{\left(\sqrt{\beta^2+z_{mn}^2}\right)}}{(\beta^2+z_{mn}^2)^2} -\frac{\left ( \left ( \beta^2-2 \right)z_{mn}^2+\beta^2+z_{mn}^4 \right)\sin{\left( \sqrt{\beta^2+z_{mn}^2} \right)}}{\left ( \beta^2+z_{mn}^2 \right)^{5/2}} \right]
\end{split}
\label{wp2}
\end{align}
\hrulefill
\vspace*{-10pt}
\end{figure*}

\noindent {\bf 3.} {\it Directivity for $v=0 \, | \, u = 1$.} The normalization integral results as in \eqref{wp3}. 
\begin{figure*}[!t]
\normalsize
\begin{align}
\begin{split}
&\mathcal{T}_{3} = \sum \limits_{n=1}^N \frac{2 A_n^2}{3} \,+\, 2 \sum\limits_{\substack{n,m=1 \\ m\neq n \\ n > m}}^N A_n A_m \cos{(\alpha_{mn})} \left[ \Psi_1 + \Psi_2 \right ], \qquad \text{where} \\
&\Psi_1 = \Biggr [ \frac{(\beta^6 (-(z_{mn}^2 - 2)) - \beta^4 (4 z_{mn}^4 + 27 z_{mn}^2 + 9) - \beta^2 z_{mn}^2 (5 z_{mn}^4 + 15 z_{mn}^2 - 72) - 2 z_{mn}^4 (z_{mn}^4 - 7 z_{mn}^2 + 12)) \sin{\sqrt{\beta^2 + z_{mn}^2}}}{(\beta^2 + z_{mn}^2)^{\frac{9}{2}}} \Biggr] \\
&\Psi_2 = \Biggr[ \frac{(\beta^6 + 3 \beta^4 (2 z_{mn}^2 + 3) - \beta^2 z_{mn}^2 (z_{mn}^2 + 72) - 6 z_{mn}^4 (z_{mn}^2 - 4)) \cos{\sqrt{\beta^2 + z_{mn}^2}}}{(\beta^2 + z_{mn}^2)^4} \Biggr ] 
\end{split}
\label{wp3}
\end{align}
\hrulefill
\vspace*{-10pt}
\end{figure*}

\noindent {\bf 4.} {\it Directivity for $v=1 \, | \, u = 1$.}
When {the exponents} $v$ and  $u$ grow, the directivity expression results increasingly complex, as depicted in \eqref{wp4} for the case $v=1; \, u = 1$.
\begin{figure*}[!t]
\normalsize
\begin{align}
\begin{split}
&\mathcal{T}_{4} = \sum \limits_{n=1}^N \frac{2 A_n^2}{15} \,+\, 2 \sum\limits_{\substack{n,m=1 \\ m\neq n \\ n > m}}^N  A_n A_m \cos{(\alpha_{mn})} \left[ \Psi_3 + \Psi_4 \right ], \qquad  \text{where}\\
&\Psi_3 = \sin{(\sqrt{\beta^2 + z_{mn}^2})} \Biggr [ \frac{-3 \beta^{10} + \beta^8 (z_{mn}^4 - 24 z_{mn}^2 - 81) + \beta^6 (5 z_{mn}^6 + 114 z_{mn}^4 + 1611 z_{mn}^2 + 225)}{(\beta^2 + z_{mn}^2)^{13/2}} \\
& \quad + \qquad  \frac{3 \beta^4 z_{mn}^2 (3 z_{mn}^6 + 82 z_{mn}^4 - 292 z_{mn}^2 - 1350) + \beta^2 z_{mn}^4 (7 z_{mn}^6 + 69 z_{mn}^4 - 2184 z_{mn}^2 + 5400)}{(\beta^2 + z_{mn}^2)^{13/2}} \\ 
& \qquad \qquad  \qquad \qquad \qquad  \qquad \qquad \qquad  \qquad + \qquad \quad \frac{2 z_{mn}^6 (z_{mn}^6 - 21 z_{mn}^4 + 192 z_{mn}^2 - 360)}{(\beta^2 + z_{mn}^2)^{13/2}} \Biggr] \\ 
&\Psi_4 = \cos{(\sqrt{\beta^2 + z_{mn}^2})} \Biggr[ \frac{6 \beta^8 (z_{mn}^2 - 1) + \beta^6 (29 z_{mn}^4 + 336 z_{mn}^2 + 225) + 6 \beta^4 z_{mn}^2 (5 z_{mn}^4 - 71 z_{mn}^2 - 675)}{(\beta^2 + z_{mn}^2)^6} \\
& \quad \quad \quad \quad \quad \quad \quad \quad \quad - \quad \quad \frac{3 \beta^2 z_{mn}^4 (z_{mn}^4 + 208 z_{mn}^2 - 1800) + 2 z_{mn}^6 (5 z_{mn}^4 - 72 z_{mn}^2 + 360) }{(\beta^2 + z_{mn}^2)^6} \Biggr ] 
\end{split}
\label{wp4}
\end{align}
\hrulefill
\vspace*{-10pt}
\end{figure*}
\subsection{Numerical Results}
Using the scenario described in Table \ref{table1}, the directivity was computed using the closed-form expression and then compared with numerical integration{;} all computations have been carried out with MATLAB (R2016a). For numerical integrations procedure, iterative method with $10^{-6}$ relative error tolerance has been adopted. Table \ref{table2} depicts the differences in the range of $\{10^{-13}; \,\, 10^{-12}\}$ attained when directivity is calculated by the expressions proposed herein and compared with the numerical integration results. 

\begin{table}[!htbp]
\caption{Values of directivity using \eqref{equation_geral} and numerical integration.} \label{table2}
\centering
    \begin{tabular}{ccccc}
	 \hline \textbf{{Exponents}} & \textbf{Directivity} & \textbf{Numerical Integration} & \textbf{Error}  \\ \hline \hline
		$v=0 \, | \, u=0$ & $7.75$ {dBi}	& $7.75$ {dBi} &  $1.77 \times 10^{-12}$ \\  
		$v=1 \, | \, u=0$ & $5.68$ {dBi}	& $5.68$ {dBi} &  $5.80 \times 10^{-13}$ \\  
		$v=0 \, | \, u=1$ & $9.18$ {dBi}	& $9.18$ {dBi} &  $2.97 \times 10^{-12}$ \\  
    $v=1 \, | \, u=1$ & $2.38$ {dBi}  & $2.38$ {dBi} &  $5.65 \times 10^{-13}$ \\  \hline
   \end{tabular}
\end{table}

\section{Conclusions}\label{sec:concl}
This work proposes a new directivity closed-form expression for arbitrary volumetric antenna arrays considering generic antenna element patterns. The parameters used in this closed-form expression are related to the order $v$ and $u$ of the element patterns, the rectangular spatial coordinate of each antenna element, the magnitude and phase excitation coefficients (complex excitation) of all elements, and the desired angles $(\theta_0, \phi_0)$. The proposed expression is appropriated to deternine directivity of antenna arrays even under the effect of mutual coupling, using less computational resources and time, due to the fact that the derived expression is exact. Notice that {exponents} $u,v$ must lie in the following regions: $v > -\frac{1}{2}$ and $u > -1$, for $u, v \in \mathbb{Z}$; however, these restrictions do not affect the generality of the element pattern, given that the superposition of terms $\sin^u(\theta)\cos^v(\theta)$, formed by the Fourier analysis, is denoted by integer {exponents}. 

Numerical results for different element patterns {($v, u$ exponents)} have demonstrated an excellent directivity compliance when comparing attained values calculated by the expressions proposed herein with the numerical integration results, considering a wide range of permitted values of {exponents} $v$ and $u$.

\appendices
\section{Integrals in $\mathcal{F}$}\label{app:A}
The two integrals in \eqref{integral_appendix1} can be expressed as the sum of the following integrals weighted by $\frac{\sin^{2u}{\theta} \cos^{2v}{\theta}}{4\pi}$:
%
\subsection{First integral $\Lambda$}
The {solution} of the first integral $\Lambda$ is straightforward, resulting:
\begin{equation}
\Lambda =  \sum \limits_{n=1}^N A_n^2 \int \limits_{0}^{2\pi}  d\phi = 2\pi \sum \limits_{n=1}^N A_n^2
\end{equation}

\subsection{Second integral $\Xi$}
Defining:
\begin{align}
	\begin{split}
	  a_{mn} =& \, x_{mn} \sin{\theta}; \qquad
		b_{mn} = \, y_{mn} \sin{\theta}; \\
		c_{mn} =& \, z_{mn} \cos{\theta} + \alpha_{mn},
	\end{split}
\end{align}
the second integral becomes
\begin{eqnarray}
\Xi =& 2 \sum\limits_{\substack{n,m=1 \\ m\neq n \\ n > m}}^N A_n A_m \int \limits_{0}^{2\pi} \cos{\left (a\cos{\phi} + b \sin{\phi} + c \right )} d\phi \notag\\
=&4\pi \sum\limits_{\substack{n,m=1 \\ m\neq n \\ n > m}}^N A_n A_m \cos{(c_{mn})} J_0 \left (\sqrt{a_{mn}^2 + b_{mn}^2} \right ) 
\end{eqnarray}
for $\sqrt{a_{mn}^2 + b_{mn}^2} \in \mathbb{R}$ and $J_0(\cdot)$ the Bessel function of the first kind.  Substituting the values of $a_{mn}$, $b_{mn}$ and $c_{mn}$, we have:
\begin{align}
\Xi = \,\,& 4\pi \sum\limits_{\substack{n,m=1 \\ m\neq n \\ n > m}}^N A_n A_m \cos{(z_{mn} \cos{\theta} + \alpha_{mn} )} \notag\\
& \quad \qquad \times \quad J_0 \left (\sin{\theta} \sqrt{x_{mn}^2 + y_{mn}^2} \right )
\label{PSII}
\end{align}
For $\sin{\theta} \sqrt{x_{mn}^2 + y_{mn}^2} \in \mathbb{R}$, the term $\sin{\theta}$ {is} always real; therefore, one can evaluate the following equality:
\begin{align}
\sqrt{x_{mn}^2 + y_{mn}^2} = k\sqrt{(r_{x_{n}}-r_{x_{m}})^2 + (r_{y_{n}}-r_{y_{m}})^2}
\end{align}
which is always real, because $r_{x_{n}}$, $r_{y_{n}}$, $r_{x_{m}}$ and $r_{y_{m}}$ are coordinates of the position of the antenna array and $k$ is the number of wave, also real; therefore, eq. \eqref{PSII} holds.
\subsection{$\mathcal{F}$ Expression}
%
Finally, one can express the integral $\mathcal{F}$ in \eqref{integral_appendix1}  as:
\begin{gather}\label{eq:Fapp}
	\begin{split}
\mathcal{F} =& \frac{1}{2} \sin^{2u}{(\theta)}\cos^{2v}{(\theta)} \left[ \Lambda + \Xi \right] \\
= &\frac{1}{2} \sin^{2u}{(\theta)}\cos^{2v}{(\theta)} \Biggr [\sum \limits_{n=1}^N A_n^2 \qquad \qquad \qquad \quad \quad \quad\\
& \quad \quad \quad + 2\sum\limits_{\substack{n,m=1 \\ m\neq n \\ n > m}}^N A_n A_m \cos{(z_{mn} \cos{\theta} + \alpha_{mn} )} \\
& \qquad \qquad \qquad \qquad \qquad  \times J_0 \left (\sin{\theta} \sqrt{x_{mn}^2 + y_{mn}^2} \right )  \Biggr]
	\end{split}
\end{gather}
%
%
\section{Integral in $\mathcal{T}$}\label{app:B}
%
{Using} the result of Appendix \ref{app:A}, eq. \eqref{eq:Fapp}, the integral in the denominator of eq. \eqref{integral_appendix2} can be expressed as the sum of the following integrals:
\begin{align}
\begin{split}
\zeta &= \frac{1}{2} \sum \limits_{n=1}^N A_n^2 \int \limits_0^{\pi}  \sin^{2u}{(\theta)}\cos^{2v}{(\theta)}\sin{\theta} d\theta \\
\xi &= 
\sum\limits_{\substack{n,m=1 \\ m\neq n \\ n > m}}^N A_n A_m \int \limits_0^{\pi}  \sin^{2u}{(\theta)}\cos^{2v}{(\theta)}  \cos{(z_{mn} \cos{\theta} + \alpha_{mn} )} \\
& \times J_0 \left (\sin{\theta} \sqrt{x_{mn}^2 + y_{mn}^2} \right ) \sin{\theta} d\theta
\end{split}
\end{align}
%
\subsection{The $\zeta$ integral}
%
The integral $\zeta$ can be calculated as:
\begin{equation}
\zeta = \frac{1}{8}\sum \limits_{n=1}^N A_n^2 \left[ (-1)^{2v}+1]\right] \cdot \mathcal{B}\left(u+1, \, v+ \frac{1}{2}\right),
\end{equation}
where $\mathcal{B}(\cdot)$ is the Beta function{;} the result above is valid and real for $u > -1$ and $v > -\frac{1}{2}$, such that $u \in \mathbb{R}$ and $v = \frac{q}{2}$, for $q \in \mathbb{Z}.$
%
\subsection{The $\xi$ integral}
%
To solve $\xi$, we can define $\beta=\sqrt{x_{mn}^2 + y_{mn}^2}$ and use the following trigonometric substitution: 
\begin{equation}
\cos(\theta) = x \rightarrow 
\begin{cases}
-\sin{\theta} d\theta = dx \\
\sin(\theta) = \sqrt{1-x^2}
\end{cases}
\label{trans1}
\end{equation}
The integral interval changes {accordingly}:
\begin{equation}
\cos(\theta) = x \rightarrow 
\begin{cases}
\theta = 0 \rightarrow x = \cos(0) = \,1 \\
\theta = \pi \rightarrow  x = \cos(\pi) = -1
\end{cases}
\label{trans2}
\end{equation}
Changing the integration order, we can write the integral as:
\begin{align}
\begin{split}
\xi =& 
\sum\limits_{\substack{n,m=1 \\ m\neq n \\ n > m}}^N A_n A_m \int \limits_1^{-1} x^{2v}(1-x^2)^u \cos{(z_{mn} x + \alpha_{mn})} \\
&\quad \times J_0 \left (\beta \sqrt{1 - x^2} \right ) dx 
\end{split}
\end{align}
Using trigonometric relations we can rewrite the expression as:
\begin{align}
\xi = 
\sum\limits_{\substack{n,m=1 \\ m\neq n \\ n > m}}^N A_n A_m \left[ \cos{(\alpha_{mn})} \rho_1 + \sin{(\alpha_{mn})} \rho_2 \right]
\end{align}
where
\begin{align}
 \rho_1 = \int \limits_{-1}^{1} x^{2v}(1-x^2)^u \cos{(z_{mn}x)} J_0 \left (\beta \sqrt{1 - x^2} \right ) dx \\
 \rho_2 = \int \limits_{-1}^{1} x^{2v}(1-x^2)^u \sin{(z_{mn}x)} J_0 \left (\beta \sqrt{1 - x^2} \right ) dx 
\end{align}
Assuming $v \in \mathbb{Z}$, we can analyze the parity of integral argument functions:
\begin{equation}
\begin{split}
		\underbrace{x^{2v}}_\text{even}\underbrace{(1-x^2)^u}_\text{even}\underbrace{\cos{(z_{mn} x)}}_\text{even} \underbrace{J_0 \Big (\beta \sqrt{1-x^2} \Big )}_\text{even} \rightarrow \text{even function} \\ 
\underbrace{x^{2v}}_\text{even}\underbrace{(1-x^2)^u}_\text{even}\underbrace{\sin{(z_{mn} x)}}_\text{odd} \underbrace{J_0 \Big (\beta \sqrt{1-x^2} \Big )}_\text{even} \rightarrow \text{odd function} 
\end{split}
\end{equation}
Therefore, for this symmetric interval we can conclude that $\rho_2 = 0$ and $\rho_1$ is doubled for the positive interval, thus we can write:
\begin{align}\label{eq:xi}
\begin{split}
\xi =& \,2 \sum\limits_{\substack{n,m=1 \\ m\neq n \\ n > m}}^N A_n A_m \cos{(\alpha_{mn})} \int \limits_{0}^{1} x^{2v}(1-x^2)^u \cos{(z_{mn} x)}\\
&\times J_0 \left (\beta \sqrt{1 - x^2} \right ) dx
\end{split}
\end{align}
Indeed, considering $u \in \mathbb{Z}$, we can rewrite \eqref{eq:xi} using the binomial coefficient expansion as:
\begin{align}
\begin{split}
\xi =&\, 2 \sum\limits_{\substack{n,m=1 \\ m\neq n \\ n > m}}^N \sum \limits_{\kappa = 0}^u (-1)^{(u+\kappa)} A_n A_m \binom{u}{\kappa} \cos{(\alpha_{mn})}\\ 
&\times \int \limits_{0}^{1} x^{2(v+u-\kappa)}\cos{(z_{mn} x)}J_0 \left (\beta \sqrt{1 - x^2} \right ) dx
\end{split}
\label{int}
\end{align}
To solve the integral above we need to transform the following integral 
\cite{gradshteyn}:
\begin{equation}
\int_{0}^{a} \cos (cx)J_0\left(b\sqrt{a^2-x^2}\right)\,\mathrm dx = \frac{\sin (a\sqrt{b^2+c^2})}{\sqrt{b^2+c^2}}, 
\label{table}
\end{equation}
for $b\geq 0$. Now, considering $a=1$, it is straightforward write:
\begin{align}
\begin{split}
\frac{\partial^{2v}}{\partial c^{2v}} \left [\int_{0}^{1} \cos (cx)J_0\left(b\sqrt{1-x^2}\right)\,\mathrm dx \right ] = \quad \quad \quad \quad \\ \quad \quad \quad \quad (-1)^v \int_{0}^{1} x^{2v} \cos (cx)J_0\left(b\sqrt{1-x^2}\right)\,\mathrm dx
\label{der_table}
\end{split}
\end{align}
Substituting \eqref{table} into \eqref{der_table} one can write:
\begin{align}
\begin{split}
\int_{0}^{1} x^{2v} \cos (cx)&J_0\left(b\sqrt{1-x^2}\right)\,\mathrm dx = \quad \quad \quad \quad \quad \quad \quad \quad \\ 
& \qquad \quad  (-1)^v  \frac{\partial^{2v}}{\partial c^{2v}} \left [\frac{\sin (\sqrt{b^2+c^2})}{\sqrt{b^2+c^2}} \right ] 
\label{great_result}
\end{split}
\end{align}
Therefore, using the result above, \eqref{int} can be written as:
\begin{align}
\begin{split}
\xi =& \,2 \, (-1)^{(v + 2u)} \sum\limits_{\substack{n,m=1 \\ m\neq n \\ n > m}}^N \sum \limits_{\kappa = 0}^u  A_n A_m \binom{u}{\kappa} \cos{(\alpha_{mn})}\\ 
& \times \frac{\partial^{2(v+u-\kappa)}}{\partial z_{mn}^{2(v+u-\kappa)}} \left [\frac{\sin (\sqrt{\beta^2+z_{mn}^2})}{\sqrt{\beta^2+z_{mn}^2}} \right ] \quad \text{for} \quad \beta\geq 0 
\end{split}
\end{align}
\noindent we have $\beta = \sqrt{(x_{mn})^2 + (y_{mn})^2}$, which is always positive because $x_{mn}$ and $y_{mn}$ are real, therefore we can write:
\begin{align}
\begin{split}
\xi =& 2 \, (-1)^{(v + 2u)} \sum\limits_{\substack{n,m=1 \\ m\neq n \\ n > m}}^N \sum \limits_{\kappa = 0}^u  A_n A_m \binom{u}{\kappa} \cos{(\alpha_{mn})}\\ 
&\times \frac{\partial^{2(v+u-\kappa)}}{\partial z_{mn}^{2(v+u-\kappa)}} \left [\frac{\sin (\sqrt{\beta^2+z_{mn}^2})}{\sqrt{\beta^2+z_{mn}^2}} \right ] 
\end{split}
\end{align}
%
\subsection{The $\mathcal{T}$ expression}
%
Finally, the $\mathcal{T}$ integral in \eqref{eq:int_} and \eqref{integral_appendix2} can be expressed as:
\begin{align}
\begin{split}
\mathcal{T} =&  \frac{1}{2} \left[\zeta + \xi \right] = \frac{1}{8}\sum \limits_{n=1}^N A_n^2 \left[ ((-1)^{2v}+1)\mathcal{B}\left (u+1,v+\frac{1}{2}\right) \right] \\
& + 2 \, (-1)^{(v + 2u)} \sum\limits_{\substack{n,m=1 \\ m\neq n \\ n > m}}^N \sum \limits_{\kappa = 0}^u A_n A_m \binom{u}{\kappa} \cos{(\alpha_{mn})} \\ 
& \quad \quad \quad \quad \quad \quad \times \frac{\partial^{2(v+u-\kappa)}}{\partial z_{mn}^{2(v+u-\kappa)}} \left [\frac{\sin (\sqrt{\beta^2+z_{mn}^2})}{\sqrt{\beta^2+z_{mn}^2}} \right ] 
\end{split}
\end{align}
\begin{align}
\begin{split}
\text{where:}\quad \qquad  & \beta	 = \sqrt{x_{mn}^2+y_{mn}^2}, \\
& u > -1, \quad v > -\frac{1}{2}, \quad \text{and}\quad u,v \in \mathbb{Z}
\end{split}
\end{align}

\newpage

\end{document}